\documentclass[preprint,prb,showpacs,preprintnumbers,amsmath,amssymb,superscriptaddress]{revtex4}

\usepackage{graphicx}
\usepackage{dcolumn}
\usepackage{bm}

\begin{document}
\draft
\title
{
Hall effect in superconducting Fe(Se$_{0.5}$Te$_{0.5}$) thin films
}

\affiliation{Central Research Institute of Electric Power Industry, 
2-6-1 Nagasaka, Yokosuka, Kanagawa 240-0196, Japan}
\affiliation{Department of Basic Science, The Univeristy of Tokyo, 
3-8-1 Komaba, Meguro-ku, Tokyo 153-8902, Japan}
\affiliation{JST, TRIP, 
Sanbancho, Chiyoda-ku, Tokyo 102-0075, Japan}

\author{I. Tsukada}
\email{ichiro@criepi.denken.or.jp}
\affiliation{Central Research Institute of Electric Power Industry, 
2-6-1 Nagasaka, Yokosuka, Kanagawa 240-0196, Japan}
\affiliation{JST, TRIP, 
Sanbancho, Chiyoda-ku, Tokyo 102-0075, Japan}
\author{M. Hanawa}
\affiliation{Central Research Institute of Electric Power Industry, 
2-6-1 Nagasaka, Yokosuka, Kanagawa 240-0196, Japan}
\affiliation{JST, TRIP, 
Sanbancho, Chiyoda-ku, Tokyo 102-0075, Japan}
\author{Seiki Komiya}
\affiliation{Central Research Institute of Electric Power Industry, 
2-6-1 Nagasaka, Yokosuka, Kanagawa 240-0196, Japan}
\affiliation{JST, TRIP, 
Sanbancho, Chiyoda-ku, Tokyo 102-0075, Japan}
\author{T. Akiike}
\affiliation{Department of Basic Science, The Univeristy of Tokyo, 
3-8-1 Komaba, Meguro-ku, Tokyo 153-8902, Japan}
\affiliation{JST, TRIP, 
Sanbancho, Chiyoda-ku, Tokyo 102-0075, Japan}
\author{R. Tanaka}
\affiliation{Department of Basic Science, The Univeristy of Tokyo, 
3-8-1 Komaba, Meguro-ku, Tokyo 153-8902, Japan}
\affiliation{JST, TRIP, 
Sanbancho, Chiyoda-ku, Tokyo 102-0075, Japan}
\author{Y. Imai}
\affiliation{Department of Basic Science, The Univeristy of Tokyo, 
3-8-1 Komaba, Meguro-ku, Tokyo 153-8902, Japan}
\affiliation{JST, TRIP, 
Sanbancho, Chiyoda-ku, Tokyo 102-0075, Japan}
\author{A. Maeda}
\affiliation{Department of Basic Science, The Univeristy of Tokyo, 
3-8-1 Komaba, Meguro-ku, Tokyo 153-8902, Japan}
\affiliation{JST, TRIP, 
Sanbancho, Chiyoda-ku, Tokyo 102-0075, Japan}

\date{\today}

\begin{abstract}
The Hall effect is investigated for eight superconducting 
Fe(Se$_{0.5}$Te$_{0.5}$) thin films grown on MgO and LaSrAlO$_4$ substrates 
with different transition temperatures ($T_c$). 
The normal Hall coefficients ($R_{\rm H}$) have positive values with magnitude 
of 1 $\sim$ 1.5 $\times$ 10$^{-3}$~cm$^3$/C at room temperature 
for the all samples. 
With decreasing temperature, we find two characteristic types of behavior 
in $R_{\rm H}$($T$) depending on $T_c$. 
For thin films with lower $T_c$ (typically $T_c$ $<$ 5~K), 
$R_{\rm H}$ start decreasing approximately below $T$ = 250~K toward 
a negative side, some of which shows sign reversal at $T$ = 50 $\sim$ 60~K, 
but turns positive toward $T$ = 0~K. 
On the other hand for the films with higher $T_c$ (typically $T_c$ $>$ 9~K), 
$R_{\rm H}$ leaves almost unchanged down to $T$ $\approx$ 100~K, 
and then starts decreasing toward a negative side. 
Around the temperatures when $R_H$ changes its sign from positive to negative, 
obvious nonlinearity is observed in the field-dependence of Hall resistance 
as to keep the low-field $R_H$ positive while the high-field $R_H$ negative. 
Thus the electronic state just above $T_c$ is characterized by $n_e$ 
(electron density) $>$ $n_h$ (hole density) with keeping $\mu_e$ $<$ $\mu_h$. 
These results suggest the dominance of electron density to the hole density 
is an essential factor for the occurence of superconductivity 
in Fe-chalcogenide superconductors. 
\end{abstract}

\maketitle


The discovery of superconductivity in iron oxypnictides 
\cite{Kamihara1} 
opened new routes for discovery of high-temperature (high-$T_c$) 
superconductors other than cuprates. 
Triggered by the discovery of LaFeAs(O$_{1-x}$F$_x$),
\cite{Kamihara1} 
a tremendous number of studies were very quickly undertaken on these materials. 
The presence of superconductivity in Fe-based binary compounds, FeSe, 
\cite{Hsu1,Mizuguchi1} 
is surprising, because the $T_c$ of a FeSe system exceeds 36~K 
under high pressure despite its simple crystal structure,
\cite{Medvedev1} 
which is comparable to MgB$_2$, another binary compound superconductor. 
Interestingly, FeSe had already been investigated in detail, but for its 
properties as a ferromagnetic semiconductors. 
According to the first investigation by Feng {\it et al.},
\cite{Feng1} 
and by several other related works,
\cite{Wu1,Liu1,Wu2} 
the transport properties of FeSe have been summarized 
in the following four characteristic features. 
First, a majority of the charge carriers at room temperature are holes, 
and the resistivity decreases with increasing hole concentrations. 
Second, spontaneous magnetization is present at room temperatures. 
Third, band calculations suggest the coexistence of hole- and electron-type 
carriers at room temperature. 
Finally, sign reversal of the Hall coefficient $R_{\rm H}$ occurs 
below 100~K when the material is doped sufficiently with holes. 
However, the relationship between these normal state electronic properties and 
the superconducting properties of these materials has not yet been clarified.

One of the features that make Fe-based superconductors exotic is its paring 
symmetry as was first proposed for LaFeAs(O$_{1-x}$F$_x$) 
by Mazin {\it et al.,} 
\cite{Mazin1} 
and Kuroki {\it et al.,}
\cite{Kuroki1} 
which requires ($\pi$, $\pi$) nesting of the Fermi surface. 
However, for the FeSe system it remains a point of argument 
whether ($\pi$, $\pi$) nesting
\cite{Subedi1,Xia1} 
or another ($\pi$, 0) nesting
\cite{Han1} 
is responsible for the formation of Cooper pairs. 
This controversy gothers our attention and lead us to an important question 
whether the Fe-pnictides and Fe-chalcogenide is essentially the same 
sumperconductor, or not. 
To resolve this it is important to quantitatively investigate the electronic 
transport properties in the relation to the superconductivity. 
Recently, we succeeded in growing superconducting Fe(Se$_{0.5}$Te$_{0.5}$) 
thin films on MgO and LaSrAlO$_4$ (LSAO) substrates,
\cite{Imai1} 
in which we found that a choice of substrates was crucial 
for the growth of Fe(Se,Te) films. 
The strong substrates dependence has been also discussed with resepect to 
an apitaxial strain in previous works on FeSe and Fe(Se,Te) thin films.
\cite{MKWu1,Kumary1,Wang1,Nie1,Han2,Bellingeri1,Si1} 
In this paper, we report the results of detailed Hall-effect measurements 
on several Fe(Se$_{0.5}$Te$_{0.5}$) thin films with different $T_c$'s, 
conducted in order to obtain deeper insight into the electronic states 
of this material, and to investigate potential routes toward a higher-$T_c$. 
We confirm the coexistence of electrons and holes first, and, in addition, 
we find 1) that there is an intimate relation between the temperature dependence 
of $R_{\rm H}$ and $T_c$, 
and 2) that the electronic state just above $T_c$ is characterized by $n_e$ 
(electron density) $>$ $n_h$ (hole density) with keeping $\mu_e$ < $\mu_h$.


Fe(Se$_{0.5}$Te$_{0.5}$) films highly oriented along the $c$ axis were grown 
by pulsed-laser deposition as described elsewhere.
\cite{Imai1} 
We prepared eight thin-film samples for the Hall-effect measurements. 
Detailed specifications of the films are summarized in Table~\ref{Table1}. 
Six films were prepared from a stoichiometric FeSe$_{0.5}$Te$_{0.5}$ sintered 
target, and two were prepared from a FeSe$_{0.5}$Te$_{0.75}$ target 
containing excess Te. 
We prepared two films simultaneously during each deposition, 
one on MgO (100) and the other on LaSrAlO$_4$ (001) in order to 
purely extract the substrate dependence. 
Samples A and D, B and E, C and F, and G and H were each grown 
in the same deposition process. 
All of the films were deposited in a six-terminal shape using a metal mask 
appropriate for both resistivity and Hall-effect measurements. 
The crystal structure was investigated by an ordinary $\theta$-2$\theta$ 
x-ray diffraction. 
Thicknesses of the films were estimated by Dektak 6M stylus profiler. 
Resistivity and Hall effect measurements were obtained using PPMS 
for temperatures down to $T$ = 2~K.



All the films shown in this paper have a highly $c$-axis 
oriented structure in common, 
while their $c$-axis length is a little bit scattered. 
As was shown in our previous paper
\cite{Imai1} 
and is summarized again in Table~\ref{Table1}, 
the $c$-axis length of the prepared films spreads out from 5.79 to 5.93~{\AA}, 
which are shorter than that reported for polycrystalline samples.
\cite{Fang1} 
Thus, we suspected that the chemical composition of the films was deviated 
largely from that of the target, 
and carried out an EDX measurement to the selected films. 
The obtained results of Se / Te ratio is shown in Table~\ref{Table1} 
for samples A, D, G, and H. 
The Se content is richer than that of Te in common, which is probably 
the main reason for the observed short $c$-axis lengths. 
This tendency is also consistent with the fact that we obtain the films with more 
Te content when using the FeSe$_{0.50}$Te$_{0.75}$ target 
(compare samples A and G, for example). 
Such a deviation of the chemical composition between targets and films 
is probably due to the difference of the vapor pressure of Se and Te, 
which become crucial in the thin-film growth process 
using high-vaccum conditions like our PLD method. 
However, it should be also emphasized that the Se / Te ratio is not 
different much between the films prepared at the same time 
(compare samples A and D, for example). 
This means that the observed differences between the films on MgO and those 
on LaSrAlO$_4$ substrates are not caused by the difference of the Se / Te 
ratio, but rather by the difference of substrate materials. 
A more comprehensive investigation of the crystalline structure 
using four-circle x-ray diffractomator and transimssion electron microscopy 
(TEM) is necessary for revealing how the substrate gives influence 
to Fe(Se$_{0.5}$Te$_{0.5}$) on it.

\begin{figure}
\begin{center}
\includegraphics*[width=150mm]{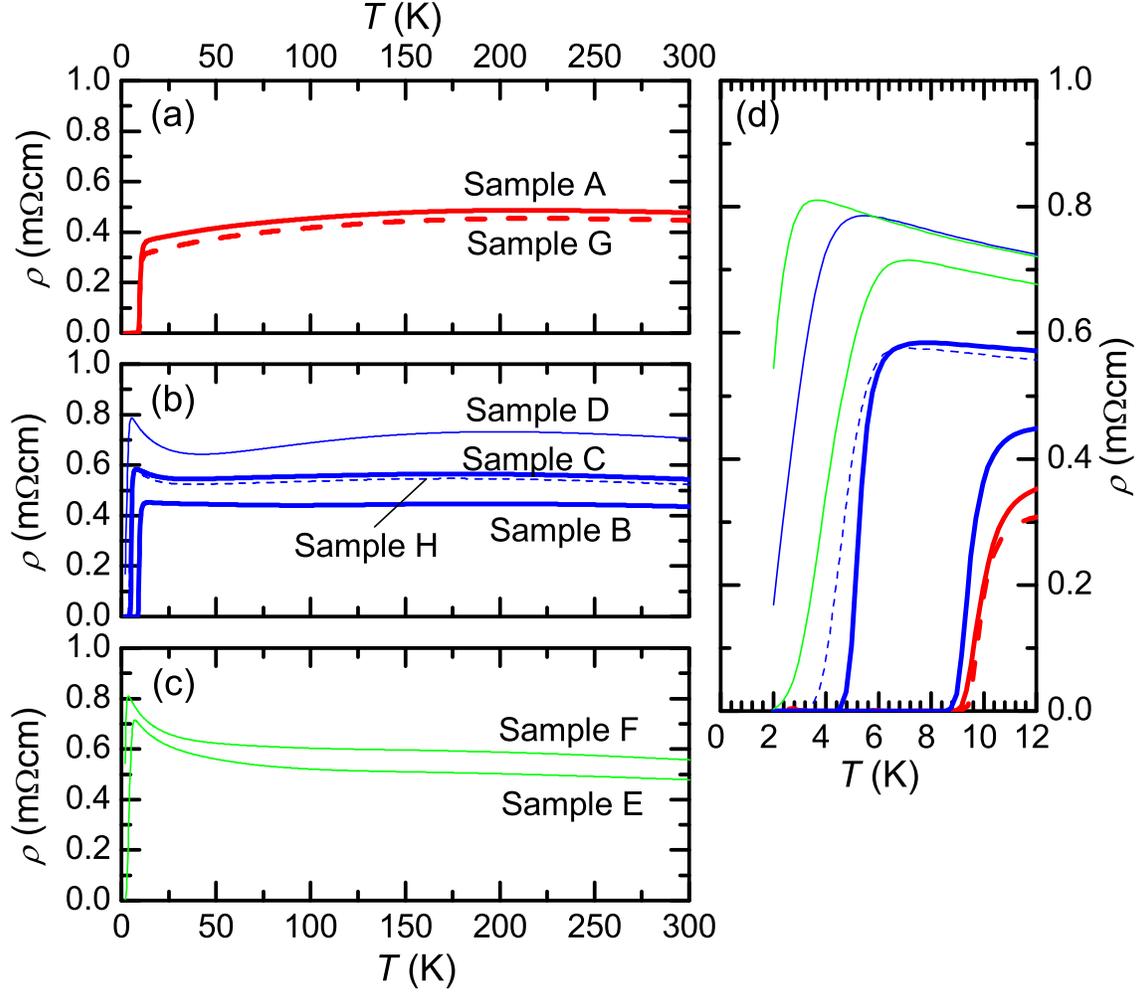}
\end{center}
\caption{(color online). 
Temperature dependence of resistivity of (a) samples A and G (red line), 
(b) samples B, C, D and H (blue line), and (c) samples E and F (green line). 
Thick and thin lines are for the films 
grown on MgO and LaSrAlO$_4$, respectively. 
(d) $\rho$($T$) around $T_c$ regions. 
}
\label{Fig1}
\end{figure}

Figure~\ref{Fig1} shows the temperature dependence of the resistivity 
for the eight samples. 
The resistivity at room temperatures are spread only within 
450 and 700~m${\Omega}$cm, which is sufficiently low for metallic 
(and superconducting) bahavior. 
However, we find that $\rho$($T$) may be classified into three groups 
from the measuremtens down to $T$ = 2~K. 
Figure~\ref{Fig1}(a) represents the most metallic group (samples A and G), 
which exhibits crossover from semiconducting (${d\rho}/dT$ $<$ 0) 
to metallic (${d\rho}/dT$ $>$ 0) behavior with decreasing temperature. 
$T_c$ exceeds 9~K, which is the highest such value among all the films. 
The second group represents an ``intermediate" group (samples B, C, D and H) 
and is illustrated in Fig.~\ref{Fig1}(b). 
This group is also characterized by ${d\rho}/dT$ $>$ 0 at intermediate 
temperatures, but the slope turns again to ${d\rho}/dT$ $<$ 0 
before the transition to superconductivity. 
The samples in this group also exhibit lower $T_c$ than those 
in the first group. 
The third group (samples E and F) is the most insulating group 
(Fig.~\ref{Fig1}(c)), 
as it never exhibits metallic (${d\rho}/dT$ $>$ 0) behavior. 
As shown in Fig.~\ref{Fig1}(d), all the films become superconducting 
even though this state is not observed for $T$ $\geq$ 2~K for samples D and F.

\begin{figure}
\begin{center}
\includegraphics*[width=150mm]{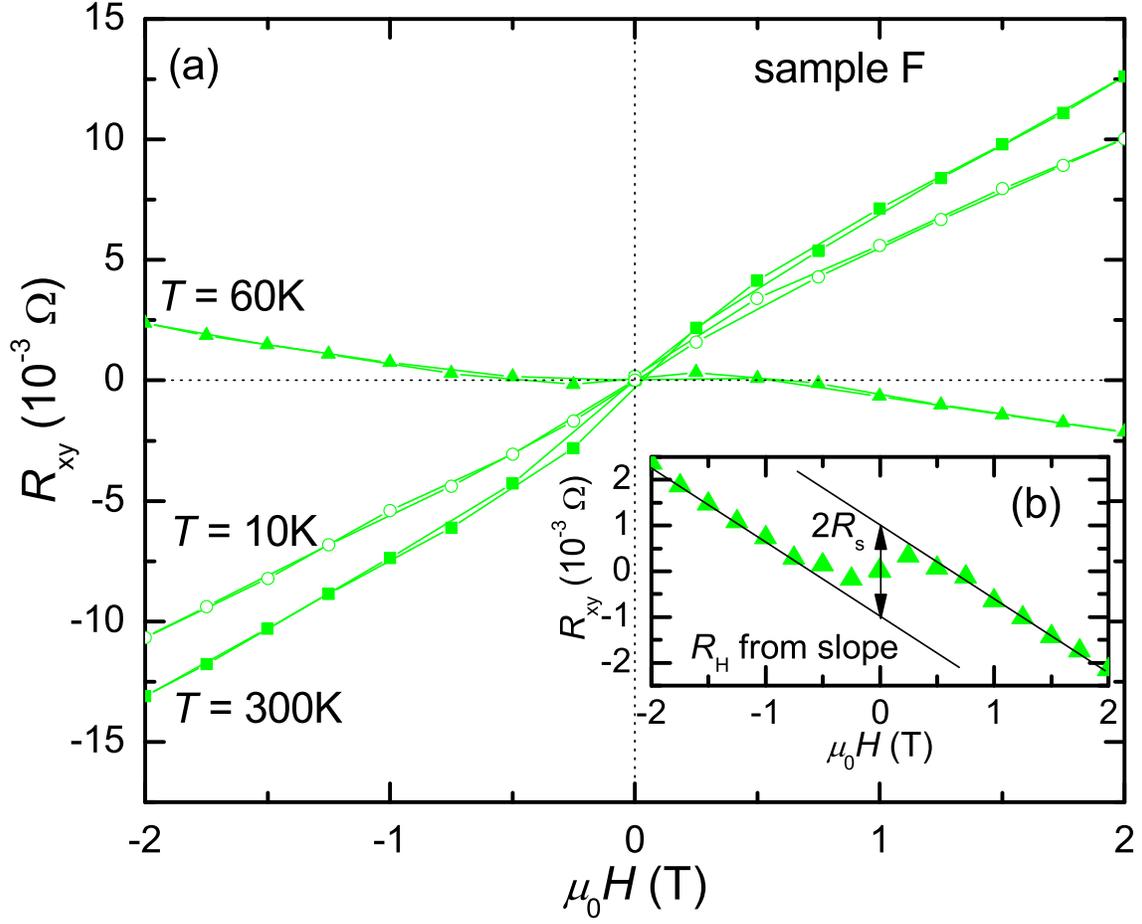}
\end{center}
\caption{(color online). 
(a) Hall resistance ($R_{xy}$) of sample F measured by sweeping field 
as $\mu_0{H}$ = 0~T $\rightarrow$ 2~T $\rightarrow$ -2~T $\rightarrow$ 0~T 
taken at $T$ = 300, 60, and 10~K. 
(b) The close-up figure of $T$ = 60~K data. 
}
\label{Fig2}
\end{figure}

For the evaluation of $R_{\rm H}$, we first need to measure transverse 
resistance $R_{xy}$ by sweeping the magnetic field, because the presence of 
the anomalous Hall effect (AHE)
\cite{AHE1}
 has already been reported for FeSe by Feng {\it et al.}
\cite{Feng1} 
An example of the typical behavior is shown for sample F in Fig.~\ref{Fig2}(a). 
We sweep the magnetic field from $\mu_0H$ = 0~T to 2~T to -2~T to 0~T, 
and observed that $R_{xy}$ showed a step-like behavior around the $H$ = 0~T 
indicating a contribution from the AHE. 
Thus, it is likely that the present Fe(Se$_{0.5}$Te$_{0.5}$) thin films 
has spontaneous magnetizaion, athough we cannot measure the magnetizaion 
due to the insufficient sample volume. 
$R_{xy}$ is expressed as the sum of a normal Hall term ($R_{\rm H}B$) 
and an anomalous Hall term ($R_s\mu_0M$). 
A steep increase in $R_{xy}$ is only observed between 
-0.5~T $<$ $\mu_0H$ $<$ 0.5~T (Fig.~\ref{Fig2}(b)), 
so we reasonably determine $R_{\rm H}$ by linearly fitting 
of $R_{xy}$ vs $\mu_0H$ between -2~T $<$ $\mu_0H$ $<$ -1~T. 
Figure~\ref{Fig2}(a) also shows that $R_{\rm H}$ $>$ 0 at $T$ = 300~K, 
that $R_{\rm H}$ $<$ 0 at $T$ = 60~K , and that $R_{\rm H}$ $>$ 0 at $T$ = 10~K. 
It should be noted that the sign of $R_s$ is not influenced by the sign 
reversals of $R_{\rm H}$. 
Although we do not further discuss the detailed $T$ dependence of $R_s$ 
in this paper, $R_s$ appears roughly proportional to $\rho$($T$) 
and/or $\rho^2$($T$), which is consistent with conventional scenarios 
of skew scattering effect and/or side jump effects.
\cite{Chien1}

\begin{figure}
\begin{center}
\includegraphics*[width=150mm]{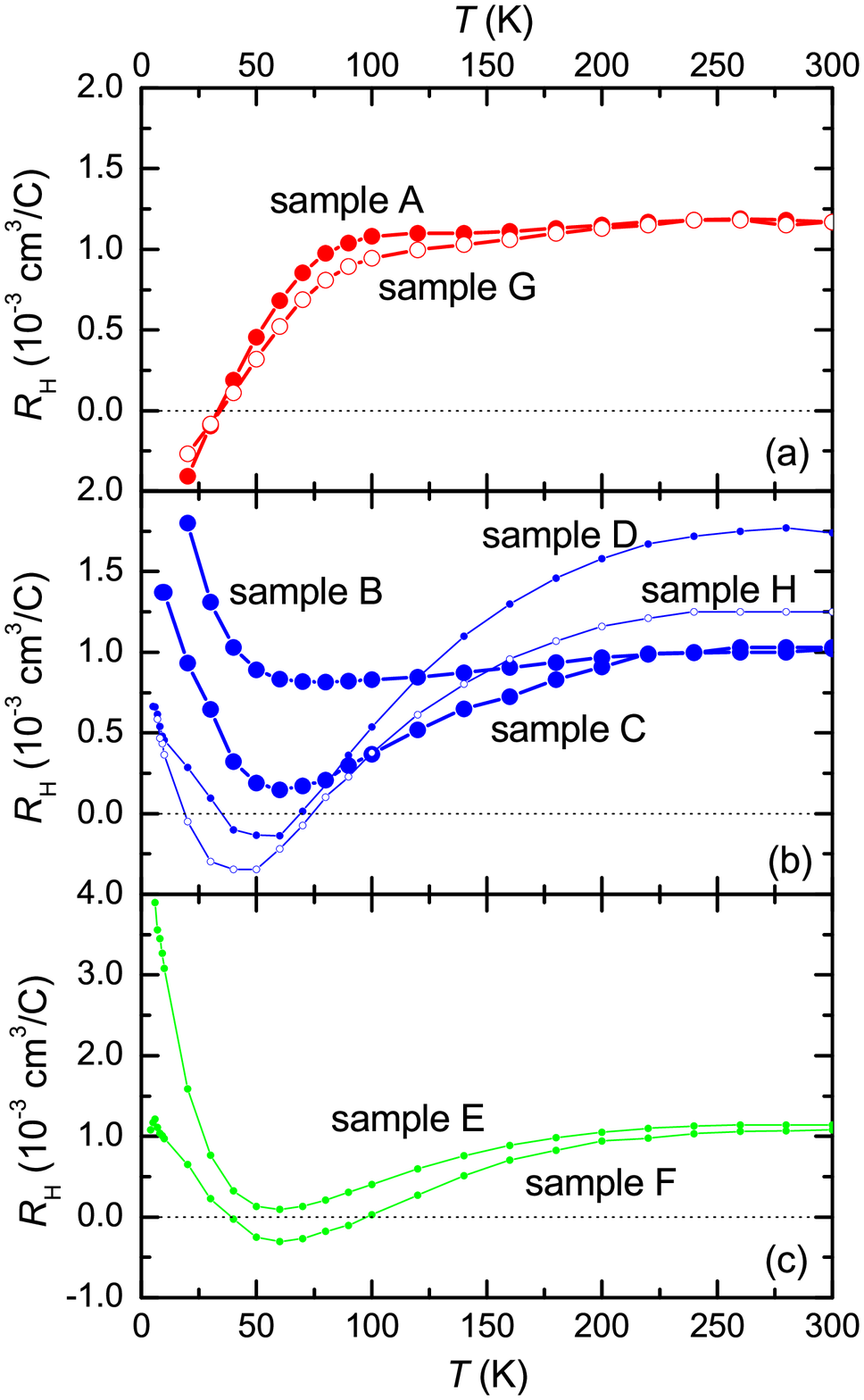}
\end{center}
\caption{(color online). 
Temperature dependence of weak-field normal Hall coefficients 
of Fe(Se$_{0.5}$Te$_{0.5}$) thin films of (a) samples A and G (red line), 
(b) samples B, C, D and H (blue line), and (c) samples E and F (green line). 
Thick and thin lines are for the film grown on MgO and LaSrAlO$_4$, 
respectively.
}
\label{Fig3}
\end{figure}

After removing the contribution associated with AHE, we plot $R_{\rm H}$ 
in the weak-field limit for all samples as shown in Fig.~\ref{Fig3}. 
$R_{\rm H}$ is almost independent of $T$ around room temperature, 
and has a value between 1 $\sim$ 1.5 $\times$ 10$^{-3}$ cm$^3$/C. 
This is consistent with the value reported by Feng and co-workers,
\cite{Feng1,Wu1} 
but it is half of the value reported by Wu {\it et al.}
\cite{MKWu1} 
The most metallic group of the samples maintains a nearly constant 
$R_{\rm H}$ down to 100~K. 
Below this temperature, $R_{\rm H}$ starts decreasing. 
We observe sign reversal only once for both samples A and G. 
It is surprising that these samples show almost identical $R_{\rm H}$($T$) 
and also $\rho$($T$), even though they were prepared from different targets. 
This suggests the existence of a close correlation between $\rho$ and $R_{\rm H}$. 
The $T$ dependence observed here is similar to that observed 
for Fe(Se$_{0.5}$Te$_{0.5}$) thin films by Wu {\it et al.},
\cite{MKWu1} 
in which the author claimed that the sign reversal is strong evidence 
of the multi-band nature of the band structure. 
Figures~\ref{Fig3}(b) and \ref{Fig3}(c) show the $R_{\rm H}$ value 
of the samples belonging to the second and the third groups, 
as a function of temperature. 
These $R_{\rm H}$ values typically become more positive at low temperatures. 
This suggests that the normal state transport properties for $T$ $\gtrapprox$ 
$T_c$ are dominated by hole-type conduction, 
which is a clear contrast to the behavior observed in samples A and G.

The temperature-dependent results suggest a finite correlation between 
the localization behavior of $\rho$($T$) and the upturn in $R_{\rm H}$. 
The samples with $T_c$ less than 5~K exhibit a gradual decrease of $R_{\rm H}$ 
with decreasing temperature starting from $T$ = 250~K. 
$R_{\rm H}$ decreases continuously until a temperature of 60~K is reached, 
it attains its minimum regardless of its sign being positive or negative, 
and finally turns increasing again to the positive side toward $T_c$. 
The upturn in $R_{\rm H}$ toward $T$ = 0~K is observed in samples C, D, E, F, 
and H while neither samples A nor G show such increase, 
which may allow us to relate the presence of this upturn 
to relatively low $T_c$. 
However, it should be noted that such an upturn toward $T_c$ is 
also observed in sample B that shows $T_c$ = 8.8~K as high as 
that of samples A and G, 
and we cannot say that $R_{\rm H}$($T$) behavior at $T$ $\gtrapprox$ $T_c$ 
and its sign just above $T_c$ is not directly related to $T_c$. 
Instead, we can find more robust correlation between $R_{\rm H}$ and $T_c$. 
Let us see $R_{\rm H}$($T$) of sample B again. 
Despite a strong upturn of $R_{\rm H}$ in sample B below 60~K, 
$R_{\rm H}$ looks almost independent of temperature between 100~K $<$ 
$T$ $<$ 250~K similar to that observed in samples A and G. 
Thus, we should conclude that a strong correlation exsists between $T_c$ 
and the $T$-dependence of $R_{\rm H}$ below 250~K.


In order to further explore the normal state transport properties 
of FeSe system, we need to explicitly treat the motion of electron- 
and hole-type carriers. 
Fortunately, there are several band calculations in the literature 
for the FeSe system, and based on these, 
it has been accepted that, at least, four bands which originate from different 
Fe $3d$ orbital cross the Fermi level.
\cite{Subedi1,Xia1} 
Two of these contribute hole-type conduction, 
and the other two contribute electron-type conduction. 
To minimize complexity, we do not consider all four of these bands, at once. 
Instead, we apply a simplified two-carrier model including one electron band 
(with electron density $n_e$ and mobility $\mu_e$) 
and one hole-type band (with hole density $n_h$ and mobility $\mu_h$),  
and from this try to extract phenomenological but 
essential behavior of $R_{\rm H}$($T$). 
Since we do not know a mathematical expression of $R_{\rm H}$ 
suitable for two-dimensional cylindrical Fermi surfaces, 
we borrow a classical expression for the Hall coefficient of three-dimensional 
isotropic semiconductors in the presence of both electron- and hole-type carriers:
\cite{Smith1}

\begin{equation}
R_H = \frac{1}{e} {\cdot} \frac{(n_h-n_eb^2)+b^2{\mu_h}^2B^2(n_h-n_e)}{(bn_e+n_h)^2+b^2{\mu_h}^2B^2(n_h-n_e)^2}, 
\label{eq1}
\end{equation}

\noindent
where $b$ = ${\mu_e} / {\mu_h}$, and $B$ is a magnetic flux density. 
This equation predicts immediately a non-linear dependence of 
$R_{xy}$ (= $R_{\rm H}\cdot B$) on applied field, 
which can be observed at field of several Tesla depending 
on the coefficients in the equation.

\begin{figure}
\begin{center}
\includegraphics*[width=150mm]{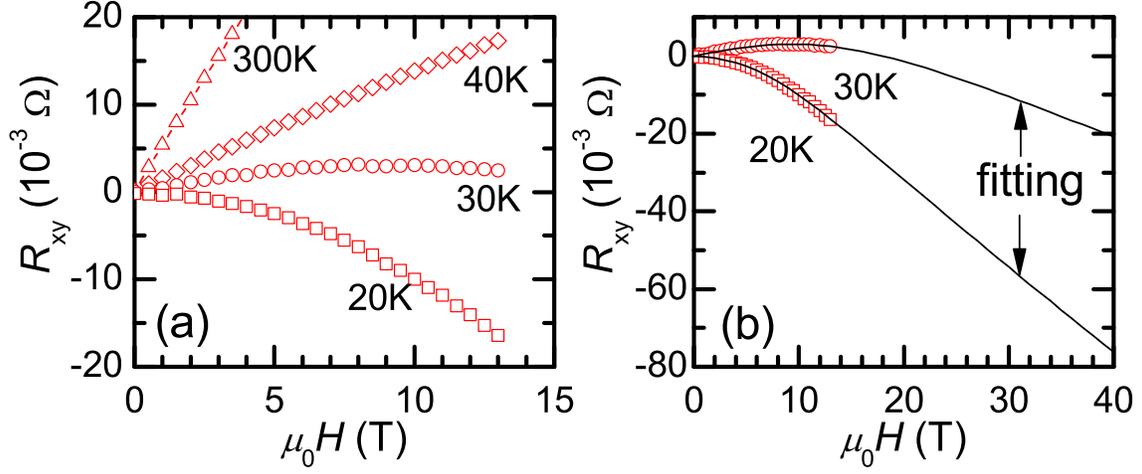}
\end{center}
\caption{(color online). 
(a) High-field $R_{xy}$ of sample G at $T$ = 300, 40, 30, and 20~K. 
The nonlinearity to $H$ becomes enhanced at $T$ = 30 and 20~K 
as a signature of coexistence of $p$ and $n$-type carriers. 
(b) Fitting with Eq.~(1) to the $T$ = 30 and 20~K data. 
}
\label{Fig4}
\end{figure}

The fitting of $R_{xy}$ is most successfully performed for sample G. 
To see how the sign change occurs in $R_{\rm H}$, 
we plot $R_{xy}$ by sweeping the field up to $\mu_0H$ = 13~T 
(Fig.~\ref{Fig4}(a)). 
Equation~(\ref{eq1}) predicts that 
$R_{\rm H}$ = $e^{-1}{\cdot}(n_h{\mu_h}^2-n_e{\mu_e}^2) / (n_e{\mu_e} + n_h{\mu_h})^2$ 
in the limit of $B$ = 0, 
while $R_{\rm H}$ = $e^{-1}{\cdot}1/(n_h-n_e)$ 
in the limit of $B$ = $\infty$. 
At $T$ = 300~K, $R_{\rm H}$ looks almost linear in $H$, indicating the hole-type 
transport is dominant. 
Furthermore, it is likely that hole-type transport is dominant even at $T$ = 40~K. 
At $T$ = 30~K and 20~K, on the other hand, 
obvious nonlinear behavior is observed. 
We successfully fit both data with Eq.~(\ref{eq1}), 
as shown by the solid lines in Fig.~\ref{Fig4}(b), 
and obtain $n_e - n_h$ = 3.38 $\times$ 10$^{22}$~cm$^{-3}$ at $T$ = 30~K, 
and $n_e - n_h$ = 1.44 $\times$ 10$^{22}$~cm$^{-3}$ at $T$ = 20~K. 
Although the values themselves are strongly dependent on the particular 
model that we used, 
in any case the density of electron-type carriers rapidly increases 
with $T$ decreasing from higher temperature, and this density 
exceeds that of hole-type carriers before superconducting transition. 
The data at $T$ = 30~K is most notable, because the slope 
at high field is negative while that at low field is positive, 
which means $p{\mu_h}^2 - n{\mu_e}^2$ $>$ 0 and $n_h - n_e$ $<$ 0. 
This gives the relation that ${\mu_h}$ $>$ ${\mu_e}$, 
which is consistent with a result previously indicated in the literature.
\cite{Wu1}


Let us first compare the present results to the reported Hall effect 
measured for single crystals. 
The normal-state Hall coefficients have been already reported for 
Ba(Fe$_{2-x}$Co$_x$As$_2$) and Ba(Fe$_2$As$_{2-x}$P$_x$) single crystals.
\cite{Albenque1,LFang1,Kasahara1} 
They commonly exhibit negative $R_{\rm H}$ at room temperatures for $x$ = 0, 
which is explained as that the electron mobility dominates the hole mobility. 
The present results for Fe(Se$_{0.5}$Te$_{0.5}$) show a clear contrast 
to Ba(Fe$_2$As$_2$); 
$R_{\rm H}$ shows a positive value exceeding 1 $\times$ 10$^{-3}$ cm$^3$/$C$ 
for the eight samples. 
By taking the fact that such positive $R_{\rm H}$ has been observed 
for Te-free FeSe thin films 
\cite{Feng1} 
and for Se-free FeTe thin films 
\cite{Han3} 
into account, 
it is likely that Fe(Se$_{1-x}$Te$_x$) has a positive $R_{\rm H}$ 
at any $x$ (0 ${\leq}$ $x$ ${\leq}$ 1), 
which is probably one of the essential differences to distinguish FeSe-based 
superconductors from FeAs-based one. 
We may consider two possible origins of the difference. 
One is that the number of outer electrons is different between 
pnictogen and chalogen, 
and the other is more specific reasons, such as band structures. 
Actually our result indicates that the hole mobility is larger than 
the electron mobility, while the opposite relation is deduced 
for Ba(Fe$_{2-x}$Co$_x$As$_2$), 
\cite{Albenque1,LFang1}
which can be attributed to the details and local features of 
the elctronic band structure.


Next, we discuss how to understand the observed variation of $T_c$ 
and the Hall coefficients in the context of reported theories. 
One of the anomalous features of Fe-based superconductors is 
their pairing symmetry.
\cite{Kuroki1} 
A likely symmerty is $s_{\pm}$ wave, 
which requires a finite nesting condition between electron- and hole-type bands 
centered at the $M$ and $\Gamma$ points in a reciprocal space, respectively. 
Singh {\it et al.} discussed how this nesting condition is influenced 
by carrier doping.
\cite{Singh1} 
They pointed out that spin density wave (SDW) instability becomes 
dominant over superconductivity when the nesting is too good, 
and with electron doping (shrinkage of the hole-type Fermi surface) 
the nesting condition becomes worse and superconductivity results. 
Our findings would be consistent with this scenario 
if the films with higher $T_c$ had smaller hole-type Fermi surfaces 
than the lower-$T_c$ films, and they do appear this way. 
The gradual decrease of $R_{\rm H}$ below $T$ = 250~K and the associated 
increase of $R_{\rm H}$ at $T$ $\gtrapprox$ $T_c$ for films with lower $T_c$ 
indicate the dominance of hole-type carriers 
throughout the whole temperature range. 
The dependence of $R_{\rm H}$ on $T$ may be due to the different 
$T$-dependences of $\mu_h$ and $\mu_e$. 
On the other hand, the higher-$T_c$ films exhibit a drastic change 
from hole- to electron-type transport, 
which strongly indicates that the Fermi level of the higher-$T_c$ films 
is located above that of the lower-$T_c$ films around $T$ $\gtrapprox$ $T_c$. 
Thus, one possible reason why sample G has a higher $T_c$ is the shrinkage 
of the hole-type Fermi surface. 
Experimentally, we confirm that the magnitude of the resistivity 
in the normal state scales well with the $c$-axis length of the films.
\cite{Imai1} 
Therefore, we infer that the short $c$-axis length and associated lattice 
deformations sensitively change the position of the Fermi level 
in Fe(Se$_{0.5}$Te$_{0.5}$) thin films, 
and that we are able to detect this shift using Hall-effect measurements. 
To confirm this inference, we need to carry out more comprehensive measurements 
of the Hall effect in FeSe systems as a function of doping and crystal structure.


In conclusion, we measure the temperature dependence of the normal 
Hall coefficients, which is positive in contrast to Ba(Fe$_2$As$_2$), 
for several Fe(Se$_{0.5}$Te$_{0.5}$) thin films 
and find a strong correlation between $R_{\rm H}$($T$) and $T_c$. 
$R_{\rm H}$($T$) of the most metallic samples remains almost constant 
down to $T$ = 100~K, and then start decreasing toward negative side, 
which is driven by the change in the population of eletron- and hole-type 
carriers, and the charge transport at temperatures just above $T_c$ 
is dominated by electron-type carriers. 
On the other hand, in more insulating samples the dominant carriers remain 
hole-type, and the $R_{\rm H}$ exhibits a different temperature dependence. 
We proposed the the analysis of non-linear Hall resistivity, 
which can be the entrance to decompose the role of carrier concentrations 
and mobilities on the electronic structure and superconductivity in the 
iron-chalcogenide superconductors.


We thank T. Kawaguchi, H. Ikuta, A. Ichinose, J. Shimoyama, and K. Kishio 
for fruitful discussions and for sharing the unpublished data. 
We also thank H. Kontani and K. Ohgushi for valuable disucssions.


\newpage

\begin{table}
\caption{Specifications of the Fe(Se$_{1-x}$Te$_x$) films.}
\begin{tabular}{ccccccc}
\hline
name & substrate & target & $T_c$ & thickness & $c$-axis length & Se / Te contents \\
     &           &        &  [K]  & [nm]      & [{\AA}]         &                  \\
\hline
sample A & MgO (100)         & FeSe$_{0.5}$Te$_{0.5}$ & 9.2   & 55  & 5.92 & 0.668 / 0.332 \\
sample B & MgO (100)         & FeSe$_{0.5}$Te$_{0.5}$ & 8.8   & 164 & 5.90 & \\
sample C & MgO (100)         & FeSe$_{0.5}$Te$_{0.5}$ & 4.6   & 135 & 5.86 & \\
sample D & LaSrAlO$_4$ (001) & FeSe$_{0.5}$Te$_{0.5}$ & $<$ 2 & 80  & 5.79 & 0.666 / 0.334 \\
sample E & LaSrAlO$_4$ (001) & FeSe$_{0.5}$Te$_{0.5}$ & 2.0   & 250 & 5.88 & \\
sample F & LaSrAlO$_4$ (001) & FeSe$_{0.5}$Te$_{0.5}$ & $<$ 2 & 190 & 5.84 & \\
\hline
sample G & MgO(100)          & FeSe$_{0.5}$Te$_{0.75}$ & 9.4  & 212 & 5.93 & 0.588 / 0.412\\ 
sample H & LaSrAlO$_4$ (001) & FeSe$_{0.5}$Te$_{0.75}$ & 3.4  & 348 & 5.90 & 0.552 / 0.448\\
\hline
\end{tabular}
\label{Table1}
\end{table}

\newpage

\newpage

\newpage

\newpage

\end{document}